# Silicon Nanowires, Catalytic Growth and Electrical Characterization

**Walter M. Weber**[*,1,3], **Georg S. Duesberg**[1,2], **Andrew P. Graham**[1], **Maik Liebau**[1], **Eugen Unger**[2], **Caroline Cheze**[1], **Lutz Geelhaar**[1], **Paolo Lugli**[3], **Henning Riechert**[2] and **Franz Kreupl**[2]

[1] Qimonda Dresden GmbH & Co., D-01099 Dresden, Germany
[2] Qimonda AG, D- 85579 Neubiberg, Germany
[3] Institute for Nanoelectronics, Technische Universität München, D-80333 Munich, Germany

**PACS** 81.16.-c, 68.65.La, 73.63.Nm, 73.21.Hb, 73.22.-f

Nominally undoped silicon nanowires (NW) were grown by catalytic chemical vapor deposition. The growth process was optimized to control the NWs diameters by using different Au catalyst thicknesses on amorphous $SiO_2$, $Si_3N_4$, or crystalline-Si substrates. For $SiO_2$ substrates an Ar plasma treatment was used to homogenize the catalyst coalescence, and thus the NWs diameter. Furthermore, planar field effect transistors (FETs) were fabricated by implementing 10 to 30 nm thin nominally undoped Si-NWs as the active region. Various silicides were investigated as Schottky-barrier source and drain contacts for the active region. For CoSi, NiSi and PdSi contacts, the FETs transfer characteristics showed *p*-type behavior. A FET consisting of a single Si-NW with 20 nanometers diameter and 2.5 µm gate-length delivers as much as 0.15 µA on-current at 1 volt bias voltage and has an on/off current ratio of $10^7$. This is in contrast to recent reports of low conductance in undoped Si.

## 1  Introduction

Recently one-dimensional semiconductor structures have attracted enormous attention. Carbon nanotubes (CNT) have shown excellent electronic properties [1,2], nevertheless the lack of control over chirality is an obstacle for future implementation in electronic applications. Silicon nanowires (NW) in contrast are always semi-conducting and easier to integrate with conventional bulk-Si technology. In this work we first present the growth optimization of nominally undoped Si-NWs. The aim is to have a better control over the diameter of Si-NWs using standard deposition methods for the catalyst. Further on, we investigate the implementation of NWs as the active region of field effect transistors (FET) with Schottky source and drain contacts.

## 2  Growth

### 2.1  Process and characterization equipment

For the catalytic growth of nominally undoped Si-NWs, sputtered Au was used as the catalyst [3]. A high precision ion-beam coater was used to reproducibly sputter the desired nominal Au thickness down to a value of 0.2 nm. The Si-NWs were grown in a 6'' wafer compatible chemical vapour deposition (CVD) chamber, of an Applied Materials P5000 cluster tool. A direct resistive heater was built in as the susceptor to achieve a good heat coupling to the substrate. On top of the chamber a downstream microwave plasma generator ($R^3T$) was mounted. This allowed a homogeneous plasma treatment over the 6'' wafer.

* Corresponding author: e-mail: walter.weber@qimonda.com, Phone: +49 89 234 44503, Fax: +49 89 234 9556966

The process gases were controlled independently by mass flow controllers. Here undiluted $SiH_4$ (monosilane) was used as the Si source; $H_2$, Ar and $N_2$ were used as the carrier and process gases. The characterization of the Si-NWs growth took place in a scanning electron microscope (SEM) Leo 1560 at 5 kV. A 200 kV high resolution transmission electron microscope (TEM) was used to image the crystal-lattice of the Si-NWs.

| Nominal Au thickness | 0.3 nm | 0.5 nm | 1 nm | 2 nm |
|---|---|---|---|---|
| NW diameter on $SiO_2$ | 13-21 nm | 20-30 nm | 25-40 nm | 30-80 nm |
| NW diameter on Si (100) | 5-12 nm | 7-18 nm | 12-25 nm | - |

**Table 1** diameter distribution of NWs. Use of Ar-plasma to achieve uniform diameters on $SiO_2$ substrates. Values are the total deviations.

## 2.2 CVD growth

As reported elsewhere [4] the size of the Au catalyst particle determines the diameter of the nanowires. Thus a homogeneous Au coalescence before growth is essential. There are various factors which determine the coagulation behaviour, such as the wetting properties and the surface-diffusion of Au on the used substrate. Also the seeding of the crystal highly depends on the substrate used, as described below.

## 2.3 Growth on amorphous substrates

As amorphous substrates 130 nm thick silane-plasma $SiO_2$ and 200 nm $Si_3N_4$ were used. First Au was sputtered with the following nominal thicknesses: 0.3, 0.5, 1 and 2nm. The annealing prior to the growth was optimized to obtain a homogeneous Au-cluster diameter distribution. All tempering processes were carried out in-situ in the CVD chamber. First, the samples were annealed at 450°C in $H_2$ atmosphere for 300 s, which lead to a rather inhomogeneous Au cluster size and consequential NW-diameter distribution. A further Ar plasma treatment for 300 s gave a homogeneous Au-particle diameter distribution as seen in Fig. 1a,b for a 0.3 nm thick Au layer. The Ar plasma heated the surface, in such a way that uniform coalescence was considerably improved. Growth then followed by maintaining the substrate temperature at 450°C and supplying a $SiH_4$ partial pressure of 5T.

For both, the $SiO_2$ and $Si_3N_4$ layers the results were similar. The different nominal Au thicknesses lead to distinct NW-diameter distributions, as listed in table 1. The data in the table show, that the thinner the sputtered Au layer is, the more uniform and smaller the NWs diameter is. Fig. 2a shows an SEM image of Si-NWs grown on top of SiO$_2$ substrates. The length of the NWs were up to 20 µm for a growth-duration of 180 s. Nanowires of up to 1/4 mm length could be grown by extending the growth duration to 30 min. It is important to notice for integration purposes, that NWs thinner than 20 nm diameter are extremely flexible as seen in Fig. 2b. TEM images revealed that the NWs with diameters between 25 and 35 nm diameter grew in the <11-2> direction as shown in Figure 3. The image shows a 35 nm thick NW with the Au/Si eutectic at the NWs tip, its *mushroom* form was observed in all of our TEM images. The radial intrusion between the eutectic and the NW is probably silicon oxide that formed as a Au-catalyzed reaction while cooling down the NWs [5]. Figure 3 also shows twin lines parallel to the growth axis ((111)-plane). This is not a general feature for growth on amorphous substrates, since mono-crystalline ones were observed as well.

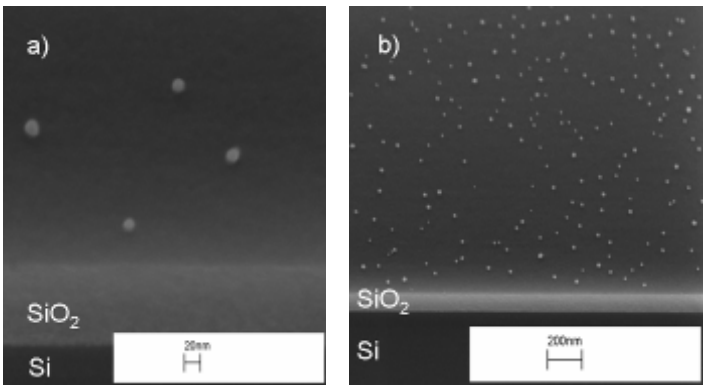

**Figure 1** SEM images of a 3 Å coalesced Au layer after annealing in $H_2$ and a posterior Ar plasma treatment. Tilted view, the substrate used is a 130 nm thick $SiO_2$ layer.

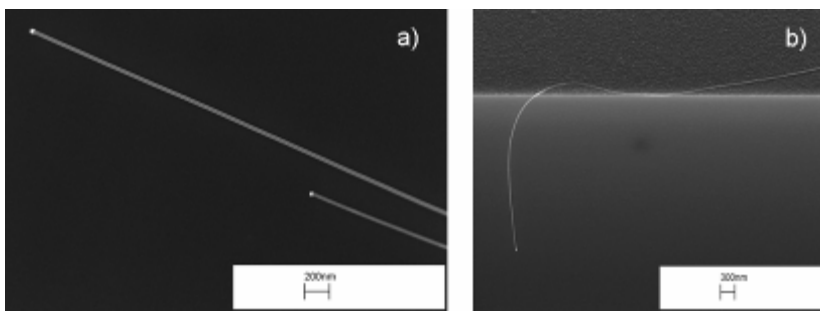

**Figure 2** Silicon NWs grown on amorphous SiO2 substrates. a) At its tip the Au/Si eutectic can be seen. b) Si-NWs longer thinner than 20 nm are extremely flexible.

## 2.4 Growth on crystalline Si-substrates.

For the growth experiments on crystalline Si, (100) oriented substrates were used. In order to seed the NWs on bare Si, the substrates surface had to be oxide-free. The removal of the native oxide and hydrogen termination of the uncovered Si surface was achieved by wet etching with hydrofluoric acid. The Au deposition and subsequent loading into the CVD chamber was carried out rapidly before re-oxidation occurred. For the Si substrates the Ar plasma treatment did not show advantages as for SiO$_2$ and Si$_3$N$_4$, thus it was excluded from the process. Growth was also performed at 450°C at the same partial pressure as described above. Typical NWs obtained from a 0.3 nm Au layer on a (100) substrate are seen in Fig. 4. From Fig. 4 a, b it is clear that the NWs grew in preferential directions, suggesting that the growth was epitaxial. Epitaxial growth has been previously reported [6,7], however, not with uniform diameter distributions using a deposited Au layer. Figure 4a shows an SEM top view, where the perpendicular NW orientation to each other is clearly seen. Therefore and since the NWs are orientated at an angle of 45° to the <100> vector of the substrates surface, Fig. 4c, it is assumed that the NWs grew in the <110>, <101>, <1-10>, <10-1> directions respectively. TEM analysis (Fig. 4d) shows the (111) planes parallel to the growth direction. This is consistent with the assumed <1-10> and <10-1> directions, since the latter are perpendicular to <111>. Notice, that the <-111> is equivalent to <111> and perpendicular to <110> and

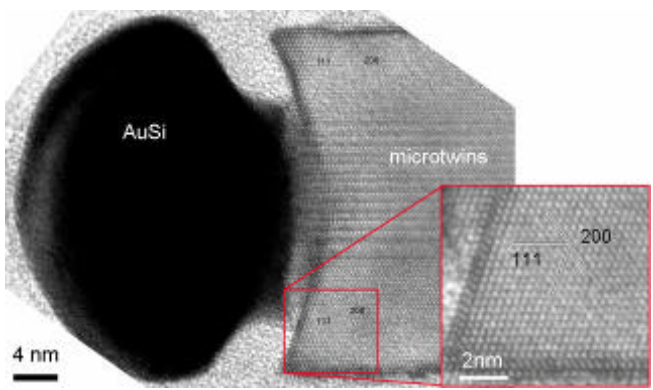

**Figure 3** HR-TEM bright-field image of the tip of a 35 nm thick Si-NW grown on SiO$_2$. It shows a <11-2> growth direction and axial twin lines

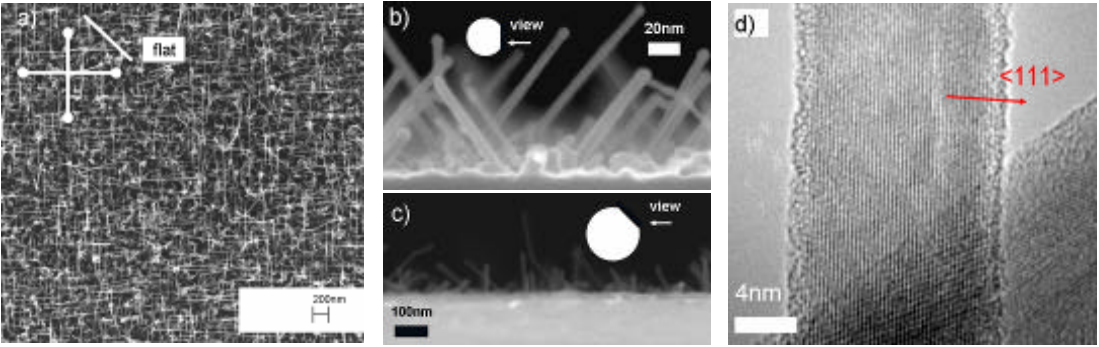

**Figure 4** Si-NWs grown on Si (100) substrates. a) SEM top view, NWs are oriented perpendicularly to each other, compare to inset diagram showing the orientations. SEM images, taken at 90° and 45° from flat (011), b) and c) respectively. The oriented growth is attributed to an epitaxial seeding process. d) TEM bright field image showing (111) planes. <1-10> could be the NWs growth direction, since it is perpendicular to the <111> direction, this is consistent with a), b) and c).

<101>. Generally, much smaller NW diameters distributions were achieved for growth on Si than on $SiO_2$ as seen in Table 1.

## 3  Electrical characterization

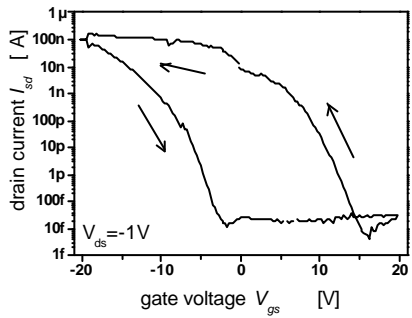

**Figure 5** Transfer characteristic of a FET consisting of a single 20 nm thick and 2.5 µm long Si-NW. Arrows show the gate voltage sweep direction. The FET displays *p*-type behaviour, a high on-current of 0.15 µA and a drain current modulation of $10^7$. Hysteresis is due to trapped surface or gate oxide charges, since the NWs are not passivated.

To investigate the electrical behaviour of these NWs, planar FETs were fabricated by implementing the undoped NWs (grown on $SiO_2$ and having diameters ranging from 13 to 30 nm) as the active regions. Here various metal silicides were used as the source and drain contacts. The test-substrates consisted of a highly *n*-doped Si wafer (100) used as a common back gate covered with a 300 nm thick $SiO_2$ gate dielectric. Metal S/D-contact structures were patterned on top with i-line photolithography and lift-off in acetone and consist of 25 nm thick titanium adhesion layer and 50nm thick cobalt, deposited by electron beam evaporation. An isopropanol solution containing the NWs, was deposited on the patterned test wafer. The silicon oxide surrounding the NWs was removed by wet etching. A successive electroless metal deposition coated the Co pads selectively, as described elsewhere [8,9]. Thus the Si-NW segments lying on the Co pads were enclosed by a thick metal layer. Co, Ni, or Pd were deposited with this method. An annealing step resulted in metal silicide formation to give Schottky contacts to the NWs at both ends. With SEM analysis single NWs bridging two contact pads were located, before electrical characterisation.

Electrical measurements were performed at room temperature with a Keithley 4200 semiconductor characterization system. A transfer characteristic of a Si-NW FET with nickel silicide source and drain contacts, a gate length ($L_g$) of 2.5 µm and a NW diameter of 20 nm is shown in Fig. 5. It shows high drain currents ($I_{ds}$) for negative gate voltages ($V_{gs}$) and low $I_{ds}$ for high positive $V_{gs}$, i.e. a clear *p*-type behaviour. Furthermore, the saturation currents are very high and reach up to 0.15 µA at -1 V drain-source bias ($V_{ds}$). This corresponds to current densities as high as 44.8 kA/cm$^2$, the highest value observed is 0.5 MA/cm$^2$ for a 20 nm diameter and 670 nm $L_g$ device. This is the highest value reported up to date for undoped Si-NWs. Fig. 5 also shows an $I_{ds}$ modulation as high as 10$^7$ through $V_{gs}$. Such high on/off $I_{ds}$ ratios are indispensable for logic applications. Another feature of our transfer characteristics is the hysteresis, which is dependent of the sweeping speed and range of $V_{gs}$. This behaviour is well known in Si-FETs and is a consequence of trapped charges at the NWs surface or at the gate oxide [10]. Passivation of Si-NW FETs with thermally grown SiO$_2$ has proven to eliminate the hysteresis [11]. The displayed unipolar *p*-type behaviour in Fig. 5 is also seen for devices with PdSi and CoSi contacts, and could be a consequence of unintentional doping during growth or processing. However, there is no evidence for this. Alternatively the unipolar behaviour can be attributed to the presence of Schottky source and drain contacts and the influence of the electric gate field on them. The Schottky barriers (SB) are lower for holes than for electrons and could in principle suppress the transport of electrons and allow thermal assisted tunnelling of holes. For bulk Si the SB-height values for electrons amount to the following values: NiSi: 0.66-0.75eV PdSi: 0.72-0.75eV and CoSi 0.64-0.68eV; for holes: NiSi: 0.39-0.48eV, PdSi: 0.39-0.42eV and CoSi 0.46-0.50eV [12,10]. Moreover for NWs the Schottky contacts are approximately one-dimensional and the SB–heights are expected to be highly reduced [13]. The band bending at the Schottky contacts imposed by the gate field can adjust the width of the SB and thus it is crucial for the transport through SiNW FETs [13]. It has been previously observed in CNT FETs, that the gate-field induced band bending at the Schottky contacts controls the current flow over the FET and not the CNT itself [14,15]. This has lead to ambipolar behaviour in CNT-FETs when using thin gate oxides and unipolar *p*-type behaviour when using thick ones as here. A similar behaviour is expected, when the CNTs are replaced by other *one*-dimensional semiconductors [14].

## 4 Summary and Conclusions

The growth process of Si-NWs has been optimized to obtain relatively homogeneous diameter distributions. For both, the crystalline Si and amorphous SiO$_2$ and Si$_3$N$_4$ substrates, thinner nominal Au layer thicknesses imply thinner NWs with smaller diameter distributions. An Ar plasma process has helped to improve uniform coalescence of the Au catalyst on SiO$_2$ and Si$_3$N$_4$ substrates. Nanowires grown on crystalline (100) Si have a clear orientation, which is attributed to the <110> and equivalent growth directions. Planar field effect transistors with undoped Si-NWs as their active region and different metal silicides as Schottky contacts, were fabricated and characterized. NiSi, CoSi and PdSi contacted NWs all show *p*-type behaviour. The current densities are as high as 0.5 MA/cm$^2$ at -1 V bias and $I_{ds}$ on/off ratio is as high as 10$^7$, these values are the highest reported up to date for nominally undoped Si-NWs and reflect the high potential of quasi-one dimensional Si structures.

**Acknowledgements** We highly appreciate the TEM imaging performed by A. Rucki and H. Cerva from Siemens.